\documentclass[twocolumn,english,pra]{revtex4-1}
\usepackage[T1]{fontenc}
\usepackage[latin9]{inputenc}
\usepackage{color}
\usepackage{amsmath}
\usepackage{amssymb}
\usepackage{graphicx}
\usepackage{esint}

\makeatletter
\@ifundefined{textcolor}{}
{%
 \definecolor{BLACK}{gray}{0}
 \definecolor{WHITE}{gray}{1}
 \definecolor{RED}{rgb}{1,0,0}
 \definecolor{GREEN}{rgb}{0,1,0}
 \definecolor{BLUE}{rgb}{0,0,1}
 \definecolor{CYAN}{cmyk}{1,0,0,0}
 \definecolor{MAGENTA}{cmyk}{0,1,0,0}
 \definecolor{YELLOW}{cmyk}{0,0,1,0}
 }

\@ifundefined{definecolor}
 {\usepackage{color}}{}
\makeatother

\makeatother

\usepackage{babel}
\begin{document}

\title{Monogamy inequalities for the EPR paradox and quantum steering }

\author{M. D. Reid}

\affiliation{Centre for Atom Optics and Ultrafast Spectroscopy, Swinburne University
of Technology, Melbourne 3122, Australia}
\begin{abstract}
Monogamy inequalities for the way bipartite EPR steering can be distributed
among $N$ systems are derived. One set of inequalities is based on
witnesses with two measurement settings, and may be used to demonstrate
correlation of outcomes between $ $two parties, that cannot be shared
with more parties. It is shown that the monogamy for steering is directional.
Two parties cannot independently demonstrate steering of a third system,
using the same two-setting steering witness, but it is possible for
one party to steer two independent systems. This result explains the
monogamy of two-setting Bell inequality violations, and the sensitivity
of the continuous variable (CV) EPR criterion to losses on the steering
party. We generalise to $m$ settings. A second type of monogamy relation
gives the quantitative amount of sharing possible, when the number
of parties is less than or equal to $m$, and takes a form similar
to the Coffman-Kundu-Wootters (CKW) relation for entanglement. The
results enable characterisation of the tripartite steering for CV
Gaussian systems and qubit GHZ and W states. 
\end{abstract}
\maketitle

\section{Introduction}

Entanglement is a major resource for quantum communication and information
processing \cite{Ent}. An important advantage of quantum communication
is the potential for an unprecedented security \cite{cry scheme benn,bellcry,bellsecu,steer ent cry}.
In quantum information, the security is based on properties, like
the no-cloning theorem \cite{noclone}, that are fundamental to quantum
mechanics, but not classical mechanics. 

A property closely connected to the no-cloning theorem is the lack
of shareability of entanglement between a number of parties. A quantitative
formulation for the way entanglement can be shared among three qubit
systems was presented by Coffman, Kundu and Wootters (CKW) \cite{ckw}
and represented an advance in understanding multipartite entanglement.
Their formulation was defined in terms of the concurrence measure
$C_{AB}$ of bipartite entanglement between two qubits, A and B \cite{conc}.
The CKW monogamy inequality is 
\begin{equation}
C_{AB}^{2}+C_{AC}^{2}\leq C_{A\{BC\}}^{2}\label{eq:CKW}
\end{equation}
where $C_{A\{BC\}}$ is the concurrence of the bi-partition $A$ with
the group $\{BC\}$. This relation illustrates that maximum entanglement
can be shared between two parties only.

Despite the importance of monogamy relations for quantum information,
the knowledge of quantitative relations for other forms of entanglement
is so far rather limited. It is known that the CKW relation can be
extended to $N$ qubits \cite{obs }, and that a violation of the
two-setting Bell inequalities \cite{Bell} is completely monogamous
\cite{bellmongisin,bell mon,bellmon2}, a property that underpins
the extra {}``device independent'' security provided by quantum
cryptography using Bell states \cite{bellsecu}. Two-party monogamy
does not however apply to Bell inequalities involving three measurement
settings per site \cite{gisinbell}. There is also a relative lack
of quantitative knowledge about the shareability of nonlocality in
the more complex continuous variable (CV) systems, although there
have been new investigations for Svetlichny\textquoteright{}s nonlocality
{[}15{]} and much progress has been made for the entanglement of CV
Gaussian states {[}16{]}, for which quantitative monogamy relations
have been worked out {[}17{]}. Recent work \cite{entform,olivmon-renyi}
analyses the reasons for the difficulty in developing monogamy relations
using more general measures of entanglement, such as entanglement
of formation. In Gaussian CV cryptography \cite{cv cry,steer ent cry},
because Bell inequalities are not directly violated \cite{ou}, the
monogamy of other forms of nonlocality is likely to be especially
useful.

The objective of this paper is to understand more about the monogamy
associated with the Einstein-Podolsky-Rosen (EPR) paradox \cite{epr,ou,rmp}.
This is the subclass of entanglement called {}``quantum steering''
\cite{Schrodinger} that was first formalised as a distinct type of
nonlocality by Wiseman, Jones and Doherty \cite{hw-steering-1,hw-np-steering-1,hw2-steering-1,smith UQ steer ,witt NJP zeilloopholefree,loopholefreesteering}.
Comparatively little is known about the shareability of this nonlocality,
which we refer to as {}``EPR steering'' \cite{EPRsteering-1}. 

Here, we will derive monogamy relations that quantify the amount of
bipartite EPR steering that can be shared by a number of parties.
As might be expected, we find the lack of shareability is greater
than for entanglement. An important feature of EPR steering monogamy
is its \emph{directionality}. While entanglement is defined symmetrically
with respect to both parties, this is not true of steering or the
EPR paradox \cite{hw-steering-1}. {}``One-way'' steering has been
realised \cite{asymmurray,kate,hand}: that party $A$ may {}``steer''
another system $B$, does not imply the converse. This property has
implications for the way EPR steering can be used to achieve secure
quantum communication \cite{steer ent cry}. In this paper, we identify
the directionality associated with steering monogamy.

Like the CKW result, the relations derived here are expressed in terms
of inequalities. The monogamy relations are specific to particular
EPR steering witnesses. We introduce two- and three- setting {}``steering
parameters'' $S_{B|A}^{(2)}$ and $S_{B|A}^{(3)}$, that involve
the variances of Pauli spin matrices, and prove monogamy relations
that apply to three qubits $A$, $B$ and $C$: 
\begin{equation}
S_{B|A}^{(2)}+S_{B|C}^{(2)}\geq2\max\{1,S_{B|\{AC\}}^{2}\}\label{eq:steerqubit-1}
\end{equation}
and $S_{B|A}^{(3)}+S_{B|C}^{(3)}+S_{B|D}^{(3)}\geq3\max\{1,S_{B|\{AC\}}^{2}\}$,
and 
\begin{equation}
S_{B|A}^{(3)}+S_{B|C}^{(3)}\geq2S_{B|\{AC]}^{(3)}\label{eq:steerqubit3}
\end{equation}
Here $S_{B|A}^{(2)},S_{B|A}^{(3)}<1$ are criteria sufficient to demonstrate
an EPR steering of system $B$, by measurements made on system $A$.
$S_{B|A}^{(2)},S_{B|A}^{(3)}\rightarrow0$ implies maximum steering.
Similar results are derived for $m$-setting steering inequalities.
These relations imply a tight monogamy: steering of a system $B$
can only be confirmed by $m-1$ other parties, using the same $m$-setting
inequality. The monogamy of Bell inequality violations is explained,
because Bell inequalities are also steering inequalities. Using a
graphical representation based on that of Plesch and Buzek \cite{entgraph},
we apply these results to depict the distribution of bipartite steering
associated with the tripartite GHZ and W states.

More fundamental are monogamy relations based on criteria for EPR
steering that are \emph{necessary and sufficient} for detecting steering.
Restricting to CV Gaussian systems, we find that such monogamy relations
are possible. EPR steering of $B$ by $A$ exists iff one can show
that a parameter $E_{B|A}$ involving conditional variances for Bob's
system (given measurements by Alice) satisfies $E_{B|A}<1$ \cite{hw-steering-1,hw2-steering-1,eprcrit}.
This is the EPR paradox criterion in which {}``elements of reality''
deduced for Bob's system show incompatibility between local realism
and the completeness of quantum mechanics \cite{eprcrit,rmp}. For
any three parties $A$, $B$ and $C$, we will see that 
\begin{equation}
E_{B|A}E_{B|C}\geq\max\{1,E_{B|\{AC\}}^{2}\}\label{eq:monoepr-2}
\end{equation}
and $E_{B|A}+E_{B|C}\geq2\max\{1,E_{B|\{AC\}}^{2}\}$. If steering
is shared between more than two Gaussian sites, then it becomes directional.
Two systems $A$ and $C$ $ $\emph{cannot} both (Gaussian) steer
a third system $B$, but we will show by example that the converse
is not true. 

A lack of robustness of the EPR criterion to losses on the steering
party, but not on the party being steered, has been noted in experiments
\cite{bow,buono}. This effect is now explained in terms of the monogamy
relation, and is seen to be a fundamental one, independent of the
mechanism of generation of the EPR fields, or the way in which losses
are implemented. This very tight form of monogamy comes about because
the witness $E_{B|A}$ is based only on two observables $-$ position
and momentum. 

We conclude the paper with a brief discussion. The steering monogamy
inequalities are likely to be useful in establishing threshold efficiency
bounds \cite{he and r} and one-sided device-independent quantum communication
security \cite{steer ent cry}.

\section{Monogamy of two-setting and CV Gaussian steering}

\subsection{CV EPR steering}

Consider the situation of three distinct and separated systems/ parties,
labelled $A$, $B$ and $C$. For each system, quadratures $X,$$P$
are defined: $X_{A}$, $P_{A}$ for system $A$, and similarly for
$B$ and $C$. We now examine the monogamy result for the two-observable
EPR criterion used to verify the CV EPR paradox \cite{ou,rmp,epr}
. 

We begin by defining a {}``steering parameter'', that enables confirmation
of the EPR paradox between two systems $A$ and $B$ \cite{eprcrit}.
The steering parameter is 
\begin{equation}
E_{B|A}=\Delta_{inf}X_{B|A}\Delta_{inf}P_{B|A}\label{eq:eprsteer}
\end{equation}
where $(\Delta_{inf}X_{B|A})^{2}$ is the variance of the conditional
distribution for the measurement $X_{B}$, given a measurement at
$A$. We normally assume that the measurement at $A$ has been optimised,
to minimise the conditional variance value. Here, $X$ and $P$ are
scaled position and momentum quadratures, so that$ $ the Heisenberg
relation for system $B$ is given by $\Delta X_{B}\Delta P_{B}\geq1$.
A confirmation of the EPR paradox \cite{eprcrit}, and quantum steering
\cite{hw-steering-1,hw2-steering-1}, is given when 
\begin{equation}
E_{B|A}<1\label{eq:eprcrit}
\end{equation}
This type of inequality is called an \textquotedblleft{}EPR steering\textquotedblright{}
or \textquotedblleft{}steering\textquotedblright{} inequality. EPR
steering inequalities based on entropic uncertainty relations have
also been derived {[}40{]}.

Following and summarising the work of Refs. \cite{hw-steering-1,hw2-steering-1,EPRsteering-1},
we will use the terminology that measurements of the system $A$ (of
Alice) {}``steer'' another system $B$ (of Bob), if some conditions
are satisfied, that imply the ensemble for $B$ has been affected
by those measurements. This relates closely to EPR's notion of {}``spooky
action-at-a-distance'', and steering is illustrated by an EPR paradox,
when Alice's inferences about Bob's system cannot be reconciled for
consistency between local realism premises and the completeness of
quantum mechanics. Steering manifests as a failure of a local hidden
variable (LHV) theory that additionally constrains Bob's local hidden
variable system to be describable as a quantum state. If the conditional
variances for $B$ are reduced, so that $E_{B|A}<1$, then this implies
a directional EPR paradox, whereby the measurements of $A$ steer
the system $B$. Throughout this paper, we abbreviate this last phrase,
to say that {}``$A$ steers $B$'', or {}``$A$ EPR steers $B$''. 

\begin{figure}[t]
\begin{centering}
\includegraphics[width=1\columnwidth]{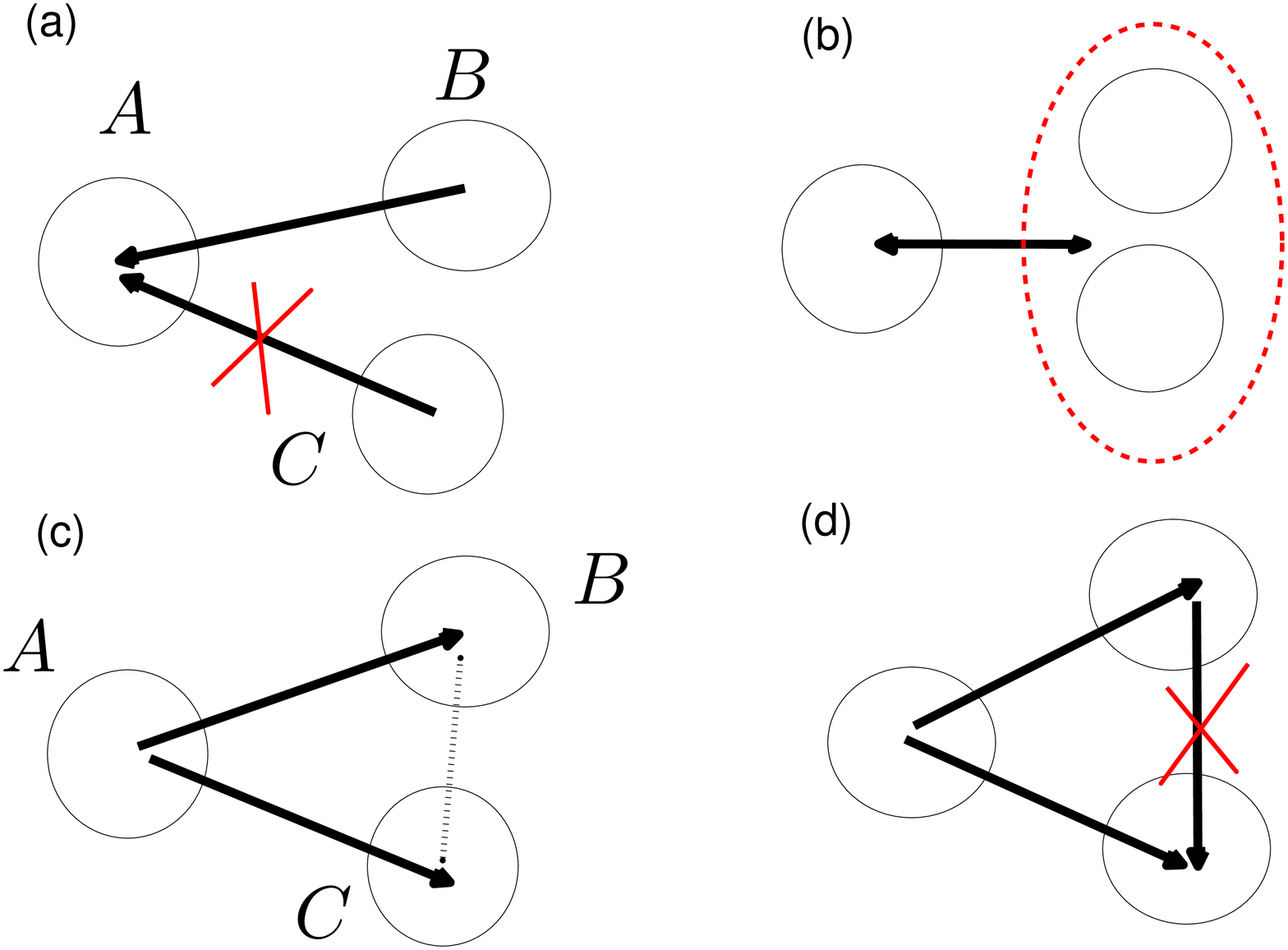}
\par\end{centering}

\caption{Depiction of ways in which bipartite steering can be shared among
three Gaussian CV systems. The results (a) and (c) also hold for steering
detected by two-setting inequalities. (a) Two parties \emph{cannot}
steer the same system. Here we depict the monogamy relation $E_{A|B}E_{A|C}\geq1$.
(b) The GHZ state has no bipartite steering between individual parties,
but there is two-way steering between any one party and the group
of the other two. (c) The dual steering by one party of two systems
can be realised. (d) The monogamy relation of (a) prevents the {}``passing
on'' of steering. If $A$ can steer $B$, and $B$ can steer $C$,
then we know that $A$ cannot steer $C$.}
\end{figure}

Now we come to the monogamy result for the CV EPR steering.

\textbf{Result (1):} If  $A$ steers $B$, in such a way that (6)
holds, then it is certain that $\Delta_{inf}X_{B|C}\Delta_{inf}P_{B|C}>1$
i.e. $C$ cannot be shown to steer $B$ by the EPR criterion. This
result is expressed as the monogamy relation 
\begin{equation}
E_{B|A}E_{B|C}\geq1\label{eq:monoepr-1}
\end{equation}
The relation has been stated and the proved in a previous paper \cite{he and r}.
We present the full details here again, because the nature of the
proof is central to the results that follow. 

\textbf{Proof:} The observer (Alice) at $A$ can make a local measurement
$O_{A}$ to infer a result for an outcome of $X_{B}$ at $B$. Denoting
the outcomes of Alice\textquoteright{}s measurement by $A_{i}$, we
can evaluate the variance of each conditional distribution $P(X_{B}|A_{i})$
and then take the average to define the inference variance   $(\Delta_{inf}X_{B|A})^{2}$.
That is, Alice's measurement is a measurement of Bob's $X_{B},$
where the uncertainty is given by $\Delta_{inf}X_{B|A}$. Similarly,
she can make a measurement of Bob's $P_{B}$, with accuracy $\Delta_{inf}P_{B|A}$.
Now, observer $C$ ({}``Charlie'') can also make inference measurements,
with uncertainty $\Delta_{inf}X_{B|C}$ and $\Delta_{inf}P_{B|C}$.
Since Alice and Charlie can make the measurements \emph{simultaneously},
it is guaranteed by the Heisenberg uncertainty relation that 
\begin{equation}
\Delta_{inf}X_{B|A}\Delta_{inf}P_{B|C}\geq1\label{eq:un}
\end{equation}
The relations are ensured because a conditional quantum density operator
$\rho_{B|\{A_{i}C_{i}\}}$ for system $B$, given the outcomes $A_{i}$
and $C_{i}$ for Alice and Charlie's measurements, can be defined,
and correctly predicts the results of all measurements. Then, the
result follows straightforwardly, on using the Cauchy-Schwarz inequality
and the definition for the inference variances given in Ref. {[}23{]}.
Similarly, Alice can measure to infer $P_{B}$ and Charlie can measure
to infer $X_{B}$, and it must also be true that 
\begin{equation}
\Delta_{inf}P_{B|A}\Delta_{inf}X_{B|C}\geq1\label{eq:un2}
\end{equation}
Hence, it follows that $E_{B|A}E_{B|C}\geq1$. $\square$

The monogamy Result (1) is depicted schematically in Figure 1a using
a generalisation of the {}``entangled graph'' representation developed
by Plesch and Buzek \cite{entgraph}. The representation depicts the
distribution of bipartite entanglement in multipartite systems. The
circles or nodes represent distinct physical systems, and a line connecting
two systems represents the bipartite entanglement between them. We
generalise the depiction in the obvious way, to denote the bipartite
steering of $A$ by $B$ by an arrow from $B$ pointing toward $A$.
We note the distinction from the graph state representation of Hein
et al \cite{graph}, in which lines between nodes represent interactions.

CV Gaussian systems are defined as those whereby the quantum states
have a positive Gaussian Wigner function and the measurements are
restricted to be Gaussian \cite{cv gauss}. For such systems, the
Result (1) is particularly useful, since in this case the optimised
EPR steering inequality (\ref{eq:eprcrit}) is \emph{necessary and
sufficient} to detect bipartite EPR steering of $B$ by $A$ \cite{hw-steering-1,hw2-steering-1}.
(The optimised inequality is that which optimises the measurement
at $A$, to minimise the conditional variances). Thus, in the Gaussian
case we can make the stronger statement, that a system can be steered
by \emph{only one other} system i.e. two distinct systems cannot independently
steer the same system (Figure 1a). We will see in Section II.C that
this sort of monogamy is one-way only. The properties of tripartite
Gaussian steering are therefore\emph{ directional}. Also, we note
that the monogamy inequality (\ref{eq:monoepr-1}) can be saturated:
$E_{B|A}=1$ was measured by Bowen et al \cite{bow} and Buono et
al \cite{buono} for a CV Gaussian state, with 50\% loss on the mode
$A$, which implies a second mode $C$ satisfying $E_{B|A}=E_{B|C}=1$.

\subsection{Monogamy of two-observable EPR steering}

The crucial aspect to the proof of the monogamy relation Result (1)
is that the steering inequality involves only two observables (measurement
{}``settings'') at each site, e.g.. position and momentum ($X$
and $P$). Similar monogamy relations can therefore be established
for other two-observable steering inequalities.

Let us consider three systems $A$, $B$ and $C$ of a fixed dimension
corresponding to a spin $J$. We define the {}``steering parameter''
\begin{equation}
S_{B|A}^{(2)}=\Bigl((\Delta_{inf}J_{B|A}^{X})^{2}+(\Delta_{inf}J_{B|A}^{Y})^{2}\Bigr)/C_{J}\label{eq:steepar}
\end{equation}
using the notation explained for (\ref{eq:eprsteer}). Here $J^{X}$,
$J^{Y}$and $J^{Z}$ are the spin components, and the constant $C_{J}$
is defined by the uncertainty relation $(\Delta J^{X})^{2}+(\Delta J^{Y})^{2}\geq C_{J}$
\cite{cj,LURhof}. EPR steering of $B$ by $A$ is confirmed when
$S_{B|A}^{(2)}<1$ \cite{high spin epr,bohm hans,multiqubits-1}.
This inequality detects what we will refer to as {}``two-observable
EPR steering'', since the inequality involves only two measurement
settings, for $J^{X}$and $J^{Y}$, at each site. 

\textbf{Result (2): }The monogamy relation 
\begin{equation}
S_{B|A}^{(2)}+S_{B|C}^{(2)}\geq2\label{eq:spnmon}
\end{equation}
holds. The proof follows as a straightforward extension of the proofs
given for Result (1) and Result (3), below.

The relation has the same consequences for monogamy as Result (1).
If EPR steering of $B$ by $A$ is confirmed by $S_{B|A}^{(2)}<1$,
then it follows that $S_{B|C}^{(2)}>1$ i.e. the system $C$ cannot
be shown to steer $A$ by using the same steering inequality.

The case $J=1/2$ is especially important, since it relates to the
original Bell states on which many experiments and quantum information
protocols are based. In terms of Pauli spin matrices $\sigma^{X}$
and $\sigma^{Y}$, we find that $S_{B|A}^{(2)}=(\Delta_{inf}\sigma_{B|A}^{X})^{2}+(\Delta_{inf}\sigma_{B|A}^{Y})^{2}$.
If bipartite EPR steering of $B$ by $A$ is observed as $\sigma_{B|A}^{(2)}=(\Delta_{inf}\sigma_{B|A}^{X})^{2}+(\Delta_{inf}\sigma_{B|A}^{Y})^{2}<1$,
then we know that for any third site $C$, there is no such steering:
that is $\sigma_{B|C}^{(2)}=(\Delta_{inf}\sigma_{B|C}^{X})^{2}+(\Delta_{inf}\sigma_{B|C}^{Y})^{2}\geq1$.
The last inequality (11) gives us information about the minimum noise
levels for Bob's qubit values as inferred by any third {}``eavesdropper''
observer at $C$, given that we know the noise elvels for Bob's qubit
values as inferred by Alice at $A$. 
\begin{figure}[t]
\begin{centering}
\includegraphics[width=1\columnwidth]{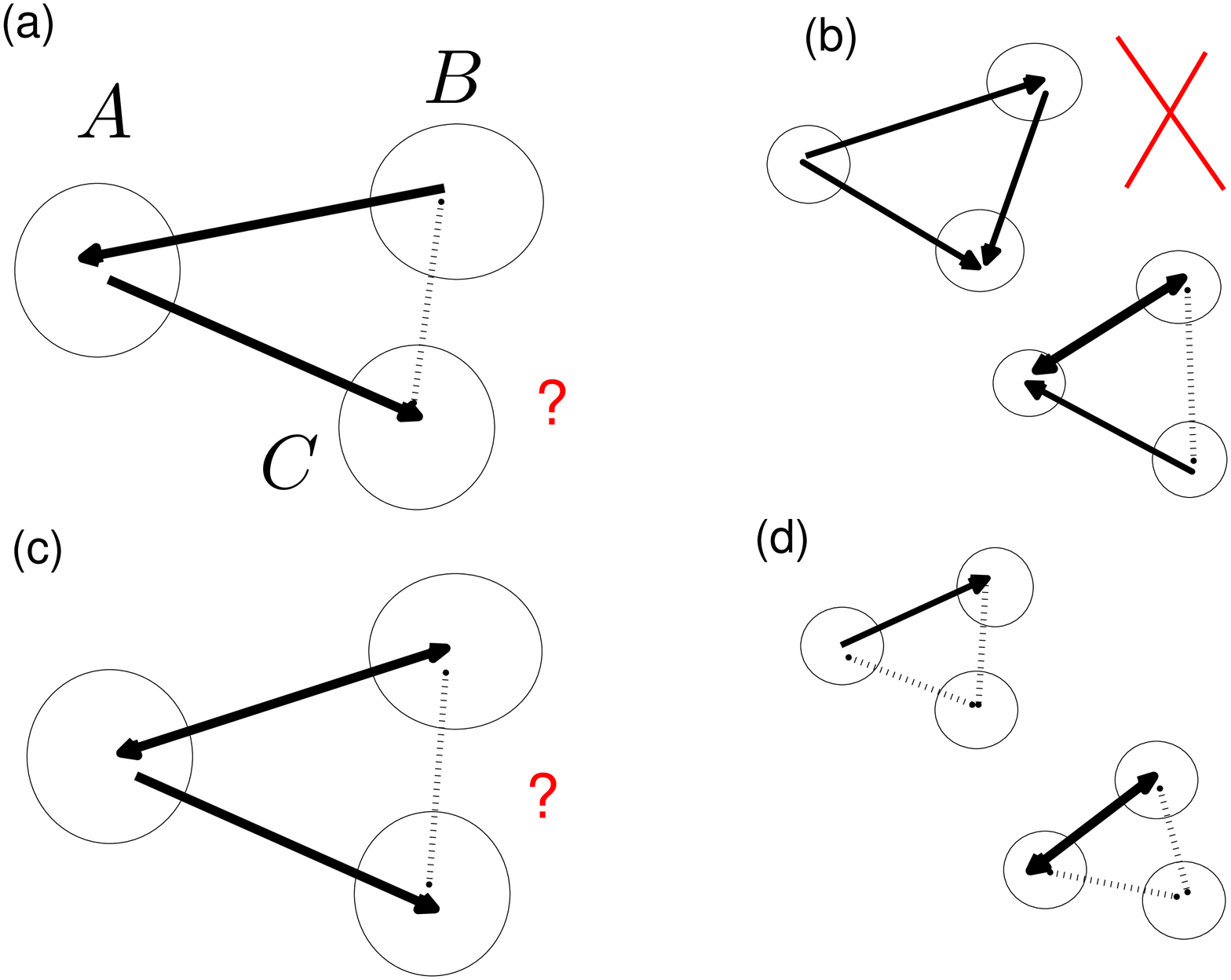}
\par\end{centering}

\caption{Other configurations for tripartite steering. Whether (a) and (c)
are possible for a Gaussian system is not established in this paper,
but the monogamy Result (1) immediately tells us that the configurations
of 1(b) are impossible, for Gaussian systems, and where steering is
detected via two-observable inequalities. The configurations (d) are
likely to be achieved by adding thermal noise to the single sites. }
\end{figure}

\subsection{Categories of tripartite Gaussian and two-observable EPR steering}

The Figures 1 and 2 show diagrammatically the possible distributions
of bipartite steering for a tripartite CV Gaussian system. The restrictions
and possibilities apply also to steering detected by a two-observable
steering inequality. These depictions are useful, because the steering,
or lack of steering, for a specific inequality can give important
information about the way correlations or noise inference levels are
shared among three parties.

Before discussing possible bipartite distributions, we recall several
properties of steering. First, steering requires entanglement. We
say the EPR steering is {}``maximum'', if the EPR conditional variances
go to zero, i.e. $E_{B|A}$, $S_{B|A}^{(2)}\rightarrow0$. For some
pure bipartite systems, the EPR steering can achieve the maximum value
and this corresponds also to the {}``strongest'' entanglement, as
measured either by concurrence \cite{conc}, or logarithmic negativity
\cite{logneg}. This is true for the two-mode squeezed state ($E_{B|A}$$\rightarrow0$)
\cite{caves sch-1} and for the qubit Bell-Bohm EPR state ($S_{B|A}^{(2)}\rightarrow0$)
\cite{Bell,bohm hans}. As not all entanglement will show EPR steering,
two systems can be entangled even if there is no EPR steering between
them.

The possibilities for steering shared between three systems are therefore
limited by the possibilities for entanglement. Two distinct types
of pure tripartite entangled qubit states exist \cite{W states}.
These are the Greenberger-Horne-Zeilinger (GHZ) \cite{ghz} and W
states. Similar states have been defined for the CV case \cite{CV GHZ,aokicv,seiji}.
Here, we discuss the bipartite distribution for specific CV Gaussian
states only, leaving the qubit case until Section VI, since for qubits
it is important to also consider steering detected by three-observable
inequalities.

The tripartite GHZ state allows no pairwise bipartite entanglement
between any of the three systems, $A$, $B$ and $C$ \cite{ckw}.
The same will be true for the EPR steering of a GHZ state (i.e.. $E_{B|A}=E_{B|C}\geq1$),
since steering is a special sort of entanglement. The GHZ state however
has bipartite entanglement between $A$ and the composite system $B-C$.
A tripartite CV GHZ state is a simultaneous eigenstate of $X_{i}-X_{j}$
($i,j=A,B,C$, $i\neq j$) and $P_{A}+P_{B}+P_{C}$ with eigenvalues
$0$ \cite{CV GHZ}. Party $A$ can choose to predict either of two
noncommuting observables (a single position and or the sum of momenta)
of the combined system $BC$, and the parties $BC$ can choose to
predict either the position or momentum of system $A$ \cite{olsen tripr,he and r}.
Thus, there is a (maximum) {}``two-way'' steering i.e.. the system
$A$ can steer the composite system $B-C$ (e.g. $E_{A|\{BC\}}=0$)
and vice versa (e.g. $E_{\{BC\}|A}=0$). This situation is depicted
in Figures 1b and 3.

Bipartite steering between two individual sites is possible for other
sorts of tripartite CV Gaussian states. However, we deduce that this
bipartite steering, in order to be consistent with the monogamy relation
Result (1), must be {}``one-way'' only. We find that the outward
{}``dual'' steering, where $A$ steers \emph{both} $B$ and $C$,
is possible (Figure 1c).  This type of tripartite steering can be
created between modes $A$, $B$, $C$ as in Figure 4. We argue as
follows. The final bipartite steering between the pair $A$ and $B$
(and similarly between $A$ and $C$) is equivalent to that between
a mode $A$ with no loss and a second mode $B$ that has been subject
to 50\% loss. That the EPR paradox (and hence steering) of the lossy
system $B$ by $A$ remains possible was summarised in Refs. \cite{rmp,asymmurray}.
The systems $B$ and $C$ are symmetric, and hence \emph{both} systems
$B$ and $C$ can be steered by $A$.

The monogamy rule Result (1)$ $ negates the possibility of the steering
{}``the other way'', that the {}``lossy'' Gaussian system $B$
(of Figure 4) steers the {}``lossless'' Gaussian system $A$. The
monogamy rule tells us that steering of $A$ by both $B$ and $C$
is ruled out. With 50\% loss on the original $B'$ channel, there
will be symmetry of the correlation between $A$ and $C$, and $A$
and $B$, in which case if $B$ can steer $A,$ then so can $C$.
This would lead to a contradiction of Result (1). That the EPR paradox
\emph{cannot} be demonstrated with 50\% loss on the steering channel
was noted experimentally \cite{bow,buono}. 

There are some open questions. The monogamy Result (1) tells us that
if $A$ can steer $B$, and $B$ can steer $C$, then $A$ \emph{cannot}
steer $C$, so that two-observable steering cannot be {}``passed
on'' (Figure 1d and 2a). It is left unaddressed however whether the
scenario of Figure 2a and c is possible, although for qubits, the
state discussed by Plesch and Buzek \cite{entgraph} will give this
possibility. The arrangements of Figure 2d are not ruled out, and
are likely to be realised by adding noise to specific sites, based
on results that indicate steering of a system $B$ by another ($A$)
is lost if thermal noise is added to $B$ \cite{HR thermal}. Another
unaddressed question concerns \emph{how} the one-way dual steering
of Figure 1c can be shared. We might expect that {}``once split''
the degree of steering would be reduced, in accordance with a monogamy
rule like that of CKW. 

\begin{figure}[t]
\begin{centering}
\includegraphics[width=1\columnwidth]{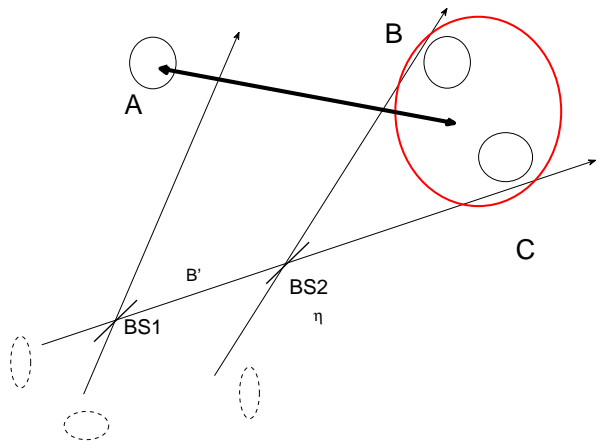}
\par\end{centering}

\caption{Schematic of the generation and EPR steering of the CV GHZ state,
which shows the tripartite steering of Figure 1c. The strong bipartite
steering and entanglement of the two-mode squeezed state can be generated
by interfering two squeezed modes at a beam splitter (BS1). }
\end{figure}

\begin{figure}[t]
\begin{centering}
\includegraphics[width=1\columnwidth]{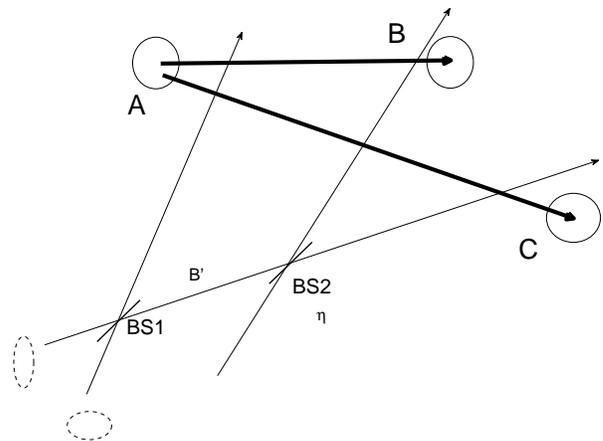}
\par\end{centering}

\caption{Schematic of the generation of the {}``dual'' EPR steering as depicted
in Figure 2c. Strong bipartite two-way EPR steering is first created
between $A$ and $B'$. The tripartite steering of Figure 2c is generated
using the second BS2 with vacuum input and efficiency of transmission
$\eta=0.5$.}
\end{figure}

\section{Multi-observable Qubit and Qudit steering monogamy relations }

More monogamy relations may be derived for EPR steering inequalities
that involve $m$ observables i.e. $m$ measurement settings, at each
site. We show that no more than $m-1$ independent parties can demonstrate
{}``steering'' of a system $B$, using the same $m$-observable
steering inequality.

\subsection{Bohm's EPR paradox monogamy}

We consider a bipartite EPR steering inequality that involves $3$
observables: $J^{X},J^{Y},J^{Z}$. We define the steering parameter
\begin{equation}
S_{B|A}^{(3)}=\Bigl((\Delta_{inf}J_{B|A}^{X})^{2}+(\Delta_{inf}J_{B|A}^{Y})^{2}+(\Delta_{inf}J_{B|A}^{Z})^{2}\Bigr)/J\label{eq:steerp}
\end{equation}
EPR steering of system $A$ by $B$ is obtained when $S_{B|A}^{(3)}<1$,
which confirms Bohm's EPR paradox for spins when $J=1/2$ \cite{EPRsteering-1,bohm hans}.
This steering inequality was derived from the uncertainty relation
$(\Delta J^{X})^{2}+(\Delta J^{Y})^{2}+(\Delta J^{Z})^{2}\geq J$
that applies to all quantum states of fixed spin $J$ i.e. to qudit
systems of dimension $d=2J+1$ \cite{LURhof}. For two qubit systems,
 $S_{B|A}^{(3)}=\bigl((\Delta_{inf}\sigma_{B|A}^{X})^{2}+(\Delta_{inf}\sigma_{B|A}^{Y})^{2}+(\Delta_{inf}\sigma_{B|A}^{Z})^{2}\bigr)/2$.

\textbf{Result (3):} We can apply the method of proof of Result (1),
to derive the monogamy steering relation. 
\begin{equation}
S_{B|A}^{(3)}+S_{B|C}^{(3)}+S_{B|D}^{(3)}\geq3\label{eq:steermon3}
\end{equation}

\textbf{Proof:} The observer at $A$ (Alice) can make the measurement
that gives her the value of Bob's observable $J_{B}^{X}$ with uncertainty
$\Delta_{inf}J_{B}^{X}$. The observer at $C$ (Charlie) can make
the measurement that gives the result for Bob's $J_{B}^{Y}$ with
uncertainty $\Delta_{inf}J_{B}^{Y}$, \emph{and} the observer at $D$
can make the measurement that gives the result for Bob's $J_{B}^{Z}$
with uncertainty $\Delta_{inf}J_{B}^{Z}$. Since the three observers
can measure simultaneously, using similar arguments as for the proof
of result 1 we see that the quantum uncertainty relation for spins
( as given above) constrains the variances to satisfy 
\begin{equation}
(\Delta_{inf}J_{B|A}^{X})^{2}+(\Delta_{inf}J_{B|C}^{Y})^{2}+(\Delta_{inf}J_{B|D}^{Z})^{2}\geq J\label{eq:1}
\end{equation}
Similarly, 
\begin{equation}
(\Delta_{inf}J_{B|A}^{Y})^{2}+(\Delta_{inf}J_{B|C}^{Z})^{2}+(\Delta_{inf}J_{B|D}^{X})^{2}\geq J\label{eq:2}
\end{equation}
and also 
\begin{equation}
(\Delta_{inf}J_{B|A}^{Z})^{2}+(\Delta_{inf}J_{B|C}^{X})^{2}+(\Delta_{inf}J_{B|D}^{Y})^{2}\geq J\label{eq:3}
\end{equation}
We then see that the monogamy relation (\ref{eq:steermon3}) follows,
upon adding the three inequalities. $\square$

The monogamy Result (3) does not exclude $2$ observers from being
able to steer $B$. However, the relation certainly prevents all $3$
observers from being able to demonstrate steering of the same system
$B$ via the violation of the $3$-observable steering inequalities
(\ref{eq:1}-\ref{eq:3}) (i.e. we can not attain $S_{B|A}^{(3)}<1$,
$S_{B|C}^{(3)}<1$ and $S_{B|D}^{(3)}<1$. We can extend the proof
of Result (3), to derive similar results involving $m$-observable
steering inequalities.\_

\subsection{Steering inequalities with $m$ observables}

Steering inequalities for two qubit systems have been derived in Refs.
\cite{EPRsteering-1,hw-np-steering-1,witt NJP zeilloopholefree,loopholefreesteering}.
The multi-observable steering inequalities derived by Saunders et
al \cite{hw-np-steering-1} and Bennet, Evans et al \cite{loopholefreesteering}
have been used in experiments that confirm steering without fair sampling
assumptions \cite{smith UQ steer ,loopholefreesteering,witt NJP zeilloopholefree}.
 Expressed in terms of correlation rather than as a noise reduction,
these steering inequalities, similar to Bell inequalities, take the
general form $\tilde{S}_{B|A}^{(m)}\leq1$, where 
\begin{equation}
\tilde{S}_{B|A}^{(m)}=\frac{1}{\mathcal{C}_{m}}\sum_{j=1}^{m}c_{j}\langle\sigma_{B}^{j}\sigma_{A}^{p_{j}}\rangle\label{eq:steersaun}
\end{equation}
Here, $\sigma_{A}^{p_{j}}$, $\sigma_{B}^{j}$ is the Pauli spin component
at angle $\theta_{p_{j}}$, $\theta_{j}$ for system $A/B$ respectively
(where $p_{j}$ is a function of $j$), $|c_{j}|=1$, $\mathcal{C}_{m}$
is a constant, and $m$ is the number of measurement settings at each
site. Steering is obtained when $\tilde{S}_{B|A}^{(m)}>1$. 

\textbf{Result (4): }The $2$-observable monogamy relation is $\tilde{S}_{B|A}^{(2)}+\tilde{S}_{B|C}^{(2)}\leq2$,
which generalises to 
\begin{equation}
\sum_{k=1}^{m}\tilde{S}_{B|A_{K}}^{(m)}\leq m\label{eq:steer2mon}
\end{equation}
where the different parties (distinct from $B$) are labelled $A_{k}$.
The result also applies to the two-observable Bell-Clauser-Horne-Shimony-Holt
(CHSH) inequality 
\begin{eqnarray*}
\tilde{S}_{B|A}^{Bell} & = & \langle\sigma_{B}^{X}\sigma_{A}^{X'}\rangle-\langle\sigma_{B}^{Y}\sigma_{A}^{Y'}\rangle+\langle\sigma_{B}^{X}\sigma_{A}^{Y'}\rangle+\langle\sigma_{B}^{Y}\sigma_{A}^{X'}\rangle\\
 & \leq & 2
\end{eqnarray*}
which is also an EPR steering inequality \cite{hw-steering-1}. EPR
steering is observed when $\tilde{S}_{B|A}^{Bell}>2$, and the monogamy
relation is $\tilde{S}_{B|A}^{Bell}+\tilde{S}_{B|C}^{Bell}\leq4$.

\textbf{Proof:} To prove (\ref{eq:steer2mon}), we recall that steering
is a failure of a special type of separable model, called a Local
Hidden State (LHS) model \cite{hw-steering-1,hw2-steering-1}. For
any LHS model 
\begin{eqnarray}
\langle\sigma_{A}^{X}\sigma_{B}^{Y}\rangle & = & \int\rho(\lambda)\langle\sigma_{A}^{X}\rangle_{\lambda}\langle\sigma_{B}^{Y}\rangle_{\lambda}d\lambda\label{eq:steerhd}
\end{eqnarray}
where $\langle\sigma_{A/B}^{X/Y}\rangle_{\lambda}$ is the predicted
average of the measurement $\sigma_{A/B}^{X/Y}$ for the local state
$\lambda$, and  the local state for the system $B$ is to be consistent
with a \emph{local quantum state} (LQS).   If the LHS model is valid,
the steering parameter can be written
\begin{eqnarray}
\tilde{S}_{B|A}^{(m)} & = & \int\rho(\lambda)\tilde{S}_{B|A}^{(m)}(\lambda)d\lambda\label{eq:steerhd-2}
\end{eqnarray}
where $\tilde{S}_{B|A}^{(m)}(\lambda)=\frac{1}{\mathcal{C}_{m}}\sum_{j=1}^{m}c_{j}\langle\sigma_{B}^{j}\rangle_{\lambda}\langle\sigma_{A}^{p_{j}}\rangle_{\lambda}$.
The steering inequality $\tilde{S}_{B|A}^{(m)}\leq1$ follows from
this assumption. A similar result holds for the Bell-CHSH inequality.

Consider an experiment where the $m$ parties $A_{1},...A_{k},..$
measure simultaneously $\sigma_{A_{1}}^{p_{1}},...\sigma_{A_{k}}^{p_{k}},..$
respectively, and the party at $B$ measures $\sigma_{B}^{j}$. We
denote the outcomes of the measurements by the symbols $\sigma_{A_{k}}^{p_{k}}$
but note they are in fact numbers, and will be identified as a {}``hidden''
variable set $\{\lambda_{1},..,\lambda_{m}\}\equiv\{\sigma_{A_{1}}^{p_{1}},...,\sigma_{A_{m}}^{p_{m}}\}$.
The state at $B$ conditioned on these outcomes is definable by a
quantum density matrix $\rho_{B|\lambda}$, and has moments (an expectation
value for $ $$\sigma_{j}^{B}$) which we once again denote by $\langle\sigma_{B}^{j}|\{\sigma_{A_{k}}^{p_{k}}\}\rangle$
(we drop the parentheses for convenience of notation). The linear
combination $\sum_{k=1}^{m}\tilde{S}_{B|A_{k}}^{(m)}$$ $can be written
in the form of an LHS model, where the probability $\rho(\lambda)$
is established as the probability $P$ of obtaining the outcomes $\{\sigma_{A_{k}}^{p_{k}}\}$
of the simultaneous measurements. Explicitly, we can write 
\begin{eqnarray}
\sum_{j=1}^{m}c_{j}\langle\sigma_{B}^{j}\sigma_{A_{j}}^{p_{j}}\rangle & = & \sum_{j}\sum_{\sigma_{A_{k}}^{p_{k}}}P(\{\sigma_{A_{k}}^{p_{k}}\})\label{eq:mat-1-1}\\
 &  & \times c_{j}\langle\sigma_{B}^{j}|\{\sigma_{A_{k}}^{p_{k}}\}\rangle\sigma_{A_{j}}^{p_{j}}\nonumber \\
\nonumber 
\end{eqnarray}
which becomes 
\begin{eqnarray}
\sum_{j=1}^{m}c_{j}\langle\sigma_{B}^{j}\sigma_{A_{j}}^{p_{j}}\rangle & = & \sum_{j}c_{j}\int\rho(\lambda)\langle\sigma_{B}^{j}\rangle_{\lambda}\langle\sigma_{A_{j}}^{p_{j}}\rangle_{\lambda}d\lambda\label{eq:tst}
\end{eqnarray}
where we see that the moments $\langle\sigma_{B}^{j}\rangle_{\lambda}$$ $
are those of the quantum state $\rho_{B|\lambda}$, and that $\langle\sigma_{A_{j}}^{p_{j}}\rangle_{\lambda}=\sigma_{A_{j}}^{p_{j}}=\lambda_{j}$.
The last line satisfies the LHS model (\ref{eq:steerhd-2}) and hence
must be less than or equal to $C_{m}$. This is true regardless of
the choice of $p_{j}$. The $\sum_{k=1}^{m}\tilde{S}_{B|A_{k}}^{(m)}$$ $
contains $m$ groups of $m$ terms like (\ref{eq:tst}), but where
different choices of simultaneous measurements are used for a given
$j$. In this way, the result follows. 

To prove the Bell-CHSH result, we consider an experiment where the
parties at $A$ and $C$ measure simultaneously $\sigma^{X'}$and
$\sigma^{Y'}$, and the party at $B$ measures $\sigma^{X}$ or $\sigma^{Y}$.
The state at $B$ conditioned on these outcomes is definable by a
quantum density matrix $\rho_{B|\lambda}$, and has moments which
we denote by $\langle\sigma_{B}^{X}|\sigma_{A}^{X'},\sigma_{C}^{Y'}\rangle$
and $\langle\sigma_{B}^{Y}|\sigma_{A}^{X'},\sigma{}_{C}^{Y'}\rangle$.
Now we see that the linear combination $\tilde{S}_{B|A}^{(Bell)}+\tilde{S}_{B|C}^{(Bell)}$$ $,
namely 
\begin{eqnarray*}
\langle\sigma_{B}^{X}\sigma_{A}^{X'}\rangle+\langle\sigma_{B}^{X}\sigma_{C}^{X'}\rangle+\langle\sigma_{B}^{Y}\sigma_{A}^{X'}\rangle+\langle\sigma_{B}^{Y}\sigma_{C}^{X'}\rangle\\
+\langle\sigma_{B}^{X}\sigma_{A}^{Y'}\rangle+\langle\sigma_{B}^{X}\sigma_{C}^{Y'}\rangle-\langle\sigma_{B}^{Y}\sigma_{A}^{Y'}\rangle-\langle\sigma_{B}^{Y}\sigma_{C}^{Y'}\rangle &  & ,
\end{eqnarray*}
can be written as consistent with a LHS model, since the probability
$\rho(\lambda)$ can be established as the probability of obtaining
the outcomes $\sigma_{A}^{X'}$ and $\sigma_{C}^{Y'}$ of the simultaneous
measurements. Explicitly, we can write 
\begin{eqnarray}
\langle\sigma_{B}^{X}\sigma_{A}^{X'}\rangle+\langle\sigma_{B}^{X}\sigma_{C}^{Y'}\rangle & = & \sum_{\sigma_{A}^{X'},\sigma_{C}^{Y'}}P(\sigma_{A}^{X'},\sigma_{C}^{Y'})\nonumber \\
 &  & \times\{\langle\sigma_{B}^{X}|\sigma_{A}^{X'},\sigma_{C}^{Y'}\rangle\sigma_{A}^{X'}\nonumber \\
 &  & \,\,\,\,\,\,\,\,\,\,\,\,\,\,+\langle\sigma_{B}^{X}|\sigma_{A}^{X'},\sigma{}_{C}^{Y'}\rangle\sigma_{C}^{Y'}\}
\end{eqnarray}
which takes the form 
\begin{eqnarray*}
\langle\sigma_{B}^{X}\sigma_{A}^{X'}\rangle+\langle\sigma_{B}^{X}\sigma_{C}^{Y'}\rangle & = & \int\rho(\lambda)\{\langle\sigma_{B}^{X}\rangle_{\lambda}\langle\sigma_{A}^{X'}\rangle_{\lambda}\\
 &  & \,\,\,\,\,\,\,+\langle\sigma_{B}^{X}\rangle_{\lambda}\langle\sigma_{C}^{Y'}\rangle_{\lambda}\}d\lambda
\end{eqnarray*}
and similarly
\begin{eqnarray*}
\langle\sigma_{B}^{Y}\sigma_{A}^{X'}\rangle-\langle\sigma_{B}^{Y}\sigma_{C}^{Y'}\rangle & = & \int\rho(\lambda)\{\langle\sigma_{B}^{Y}\rangle_{\lambda}\langle\sigma_{A}^{X'}\rangle_{\lambda}\\
 &  & \,\,\,\,\,\,\,-\langle\sigma_{B}^{Y}\rangle_{\lambda}\langle\sigma_{C}^{Y'}\rangle_{\lambda}\}d\lambda
\end{eqnarray*}
where we see that the moments $\langle\sigma_{B}^{X}\rangle_{\lambda}$,
$ $$\langle\sigma_{B}^{Y}\rangle_{\lambda}$ are those of the quantum
state $\rho_{B|\lambda}$, and $\langle\sigma_{A}^{X'}\rangle_{\lambda}=\sigma_{A}^{X'}=\lambda_{1}$
and $\langle\sigma_{C}^{Y'}\rangle_{\lambda}=\sigma_{C}^{Y'}=\lambda_{2}$.
In this way, we can write
\begin{eqnarray}
\langle\sigma_{B}^{X}\sigma_{A}^{X'}\rangle+\langle\sigma_{B}^{X}\sigma_{C}^{Y'}\rangle+\langle\sigma_{B}^{Y}\sigma_{A}^{X'}\rangle-\langle\sigma_{B}^{Y}\sigma_{C}^{Y'}\rangle\nonumber \\
=\int\rho(\lambda)\{\langle\sigma_{B}^{X}\rangle_{\lambda}\langle\sigma_{A}^{X'}\rangle_{\lambda}+\langle\sigma_{B}^{X}\rangle_{\lambda}\langle\sigma_{C}^{Y'}\rangle_{\lambda}\nonumber \\
+\langle\sigma_{B}^{Y}\rangle_{\lambda}\langle\sigma_{A}^{X'}\rangle_{\lambda}-\langle\sigma_{B}^{Y}\rangle_{\lambda}\langle\sigma_{C}^{Y'}\rangle_{\lambda}\}d\lambda\nonumber \\
=\int\rho(\lambda_{1},\lambda_{2})\{\langle\sigma_{B}^{X}\rangle_{\lambda}\lambda_{1}+\langle\sigma_{B}^{X}\rangle_{\lambda}\lambda_{2}\nonumber \\
+\langle\sigma_{B}^{Y}\rangle_{\lambda}\lambda_{1}-\langle\sigma_{B}^{Y}\rangle_{\lambda}\lambda_{2}\}d\lambda
\end{eqnarray}
The last line satisfies the LHS model (\ref{eq:steerhd-2}), on letting
$\langle\sigma_{A}^{X'}\rangle_{\lambda}=\lambda_{1}$ and $\langle\sigma_{C}^{Y'}\rangle_{\lambda}=\lambda_{2}$,
and hence must be less than or equal to $2$. By the same argument,
we can show $\langle\sigma_{B}^{X}\sigma_{A}^{Y'}\rangle+\langle\sigma_{B}^{X}\sigma_{C}^{X'}\rangle-\langle\sigma_{B}^{Y}\sigma_{A}^{Y'}\rangle+\langle\sigma_{B}^{Y}\sigma_{C}^{X'}\rangle\leq2$.
Hence, $\tilde{S}_{B|A}^{(Bell)}+\tilde{S}_{B|C}^{(Bell)}\leq4$.
$\square$

\subsection{Monogamy of steering using Bell-CHSH moments}

Two useful EPR steering inequalities that apply to the Bell Clauser-Horne-Shimony-Holt
(CHSH) state and experiment are 
\begin{equation}
\langle\sigma_{B}^{X}\sigma_{A}^{X'}\rangle-\langle\sigma_{B}^{Y}\sigma_{A}^{Y'}\rangle\leq\sqrt{2}\label{eq:eprbi}
\end{equation}
and $\langle\sigma_{B}^{X}\sigma_{A}^{Y'}\rangle+\langle\sigma_{B}^{Y}\sigma_{A}^{X'}\rangle\leq\sqrt{2}$
\cite{EPRsteering-1,hw-np-steering-1,multiqubits-1}. If either of
these inequalities is violated, steering is confirmed. Result (4)
allows us to immediately write down monogamy relations associated
with these steering inequalities: 
\begin{equation}
\langle\sigma_{B}^{X}\sigma_{A}^{X'}\rangle-\langle\sigma_{B}^{Y}\sigma_{A}^{Y'}\rangle+\langle\sigma_{B}^{X}\sigma_{C}^{X'}\rangle-\langle\sigma_{B}^{Y}\sigma_{C}^{Y'}\rangle\leq2\sqrt{2}\label{eq:monog5}
\end{equation}
and $\langle\sigma_{B}^{X}\sigma_{A}^{Y'}\rangle+\langle\sigma_{B}^{Y}\sigma_{A}^{X'}\rangle+\langle\sigma_{B}^{X}\sigma_{C}^{Y'}\rangle+\langle\sigma_{B}^{Y}\sigma_{C}^{X'}\rangle\leq2\sqrt{2}$\textbf{}.

\section{CHSH-Bell nonlocality monogamy}

Since the Bell inequalities are also steering inequalities, the monogamy
of steering implies the monogamy of the two-setting CHSH Bell inequalities.
 The CHSH Bell inequalities are 
\begin{eqnarray}
\tilde{S}_{B|A}^{Bell} & = & \langle\sigma_{B}^{X}\sigma_{A}^{X'}\rangle-\langle\sigma_{B}^{Y}\sigma_{A}^{Y'}\rangle+\langle\sigma_{B}^{X}\sigma_{A}^{Y'}\rangle+\langle\sigma_{B}^{Y}\sigma_{A}^{X'}\rangle\nonumber \\
 & \leq & 2\label{eq:bellchsh}
\end{eqnarray}
Any Bell inequality is also an EPR steering inequality \cite{hw-steering-1}.
Using Result (4), we can therefore deduce the monogamy relation for
the CHSH Bell inequality:
\begin{equation}
\tilde{S}_{B|A}^{Bell}+\tilde{S}_{B|C}^{Bell}\leq4\label{eq:sbell}
\end{equation}
The symmetry of the Bell-CHSH inequalities implies that in an experiment
(where there is a fixed choice of measurement settings at each location)
$\tilde{S}_{B|A}^{Bell}>2$ is equivalent to $\tilde{S}_{A|B}^{Bell}>2$.
That is, as indeed must be true generally given the definition of
Local hidden Variable theories {[}25{]}, the violation of a Bell inequality
implies \textquotedblleft{}two-way\textquotedblright{} steering. The
monogamy relations   $\tilde{S}_{A|B}^{Bell}+\tilde{S}_{A|C}^{Bell}\leq4$
and $\tilde{S}_{C|A}^{Bell}+\tilde{S}_{C|B}^{Bell}\leq4$ also hold.
If two parties $A$-$B$ can violate the Bell-CHSH inequality, then
the pairs $A$-$C$, and $C$-$B$ cannot.

The result for the monogamy of Bell-CHSH violations result is not
new \cite{bell mon,bellmon2,bellmongisin}. What we have discovered
from our analysis is that the monogamy follows \emph{as a result of
steering monogamy}. All two-observable (setting) steering inequalities
possess one-way monogamy. Since the Bell-CHSH violations imply two-way
steering, this is enough to explain Bell-CHSH monogamy.

Our results explain the shareability, with respect to three sites,
of the three-observable Bell inequality violation of Collins and Gisin
\cite{gisinbell}. Being three-observable steering inequalities, we
can expect however, using Result (4), that these violations cannot\emph{
}be shared among \emph{four} sites.

\section{sharing of bipartite steering}

We have seen that a very tight monogamy arises for the correlations
of a witness when the number of parties equals or exceeds $m+1$ where
$m$ is the number of observables that need to be measured at each
site. Now, we examine the constraints on the distribution of bipartite
steering, when the number of systems is less than $m+1$. 

In this Section, we therefore derive relations for steering monogamy
that are similar to the CKW inequalities, for particular witnesses.
We quantify how the {}``total amount of steering'' is shared among
the subgroups. Similar to the result for sharing of entanglement with
qubits, we find that the strongest steering exists only when all the
steering is shared between two parties. Once steering is distributed
over a series of systems, the pairwise steering will diminish. In
this paper, we prove such a rule for steering in one direction\emph{
}only.

\subsection{CV bipartite sharing }

We begin with the CV EPR steering relation (\ref{eq:eprsteer}). Given
the definition of the steering parameter $E_{B|A}$, it must be true
that
\begin{equation}
E_{B|\{AC\}}\leq E_{B|A}\label{eq:steerbi}
\end{equation}
This simple result follows, because $E_{B|A}$ is the lowest variance
product possible, that arises from the best inference of Bob's $X_{B}$
or $P_{B}$ by the group $A$ of Alice. Alice can use any local observable,
defined as a measurement performed on the system $A$. The inference
of Bob's measurement by the group $AC$, which includes both $A$
and $C$, must be at least as good as that of $A$ alone, since the
observables of system $A$ are a subset of those of the combined system
$AC$. The steering of $B$ by a combined group cannot be less effective
than that of a subset. It is also true that $E_{B|\{AC\}}\leq E_{B|C}$.
On multiplying the two inequalities together, we can easily derive
several new monogamy relations.

\textbf{Result (5): }For the three systems $A$, $B$ and $C$, it
follows that\textbf{ }
\begin{equation}
E_{B|A}E_{B|C}\geq E_{B|\{AC\}}^{2}\label{eq:steerbi2}
\end{equation}
We can express the product relation in terms of a sum relation, similar
to CKW, by using the simple identity $x^{2}+y^{2}\geq2xy$. 

\textbf{Result (6): }It is also true that\textbf{ }
\begin{equation}
E_{B|A}+E_{B|C}\geq2E_{B|\{AC\}}\label{eq:CKW-1-1-1}
\end{equation}
This follows, since we can let $x=\sqrt{E_{B|A}}$ and $y=\sqrt{E_{B|C}}$,
and use that $E_{B|\{AC\}}\leq\sqrt{E_{B|A}}\sqrt{E_{B|C}}$. Since
the maximal steering is defined when $E_{B|A}=0$, and $E_{B|A}=1$
is the threshold for steering, the direction of the inequality is
reversed as compared to that for the CKW relation. We note also that
we could choose $x=E_{B|A}$ and $y=E_{B|C}$ from which we derive
the monogamy result: 
\begin{equation}
E_{B|A}^{2}+E_{B|C}^{2}\geq2E_{B|\{AC\}}^{2}\label{eq:steersum}
\end{equation}

The relations (\ref{eq:steerbi2}-\ref{eq:steersum}) express a type
of conservation law for steering. If there is steering of $B$ by
a group $AC$ that has components $A$ and $C$, so that $E_{B|\{AC\}}<1$,
then the steering is shared among the components. The individual steering
of $B$ by $A$, or $B$ by $C$, is reduced and bounded by the monogamy
relations. 

If the property (\ref{eq:steerbi}) is specified as a condition for
a witness for EPR steering, then the relation holds for all such witnesses.
The monogamy relations (\ref{eq:steerbi2}-\ref{eq:CKW-1-1-1}) would
then become fundamental results for steering monogamy, that are non-specific
to a particular steering witness or uncertainty relation. 

The monogamy relation of Result (1) is stronger than Result (5) when
steering is present, since steering requires $E_{B|A}<1$. We thus
write the monogamy relation for the CV EPR witness (\ref{eq:eprsteer})
as 
\begin{equation}
E_{B|A}E_{B|C}\geq\max\{1,E_{B|\{AC\}}^{2}\}\label{eq:fullmon}
\end{equation}
One could test this relation experimentally, by adding noise to mode
$B$ so that $E_{B|\{AC\}}>$1. We have not demonstrated saturation
of the inequality, except where $E_{B|\{AC\}}=1$, which was discussed
in Section II.

\subsection{Qubits and qudits}

The qubit case is more interesting. Following the same approach, we
can deduce that $S_{B|A}^{(2)}\geq S_{B|\{AC\}}^{(2)}$ and $S_{B|C}^{(2)}\geq S_{B|\{AC\}}^{(2)}$
which implies 
\begin{equation}
S_{B|A}^{(2)}+S_{B|C}^{(2)}\geq\max\{2,2S_{B|\{AC\}}^{(2)}\}\label{eq:s23}
\end{equation}
and similarly 
\begin{equation}
S_{B|A}^{(3)}+S_{B|C}^{(3)}+S_{B|D}^{(3)}\geq\max\{3,3S_{B|\{ACD\}}^{(3)}\}\label{eq:s13}
\end{equation}
Also,
\begin{equation}
S_{B|A}^{(3)}+S_{B|C}^{(3)}\geq2S_{B|\{AC\}}^{(3)}\label{eq:smo}
\end{equation}
The relation (\ref{eq:smo}) for sharing of steering is significant
for qubit systems, since it will apply to limit the steering detected
using three-observable steering inequalities, for tripartite systems
(here, the number of sites is less than $m+1$). This relation resembles
the CKW relation for entanglement. We use the relation (\ref{eq:smo})
in the next Section, to derive the steering properties of the tripartite
qubit $W$ state.

\section{Steering monogamy of tripartite GHZ and W states}

\begin{figure}[t]
\begin{centering}
\includegraphics[width=1\columnwidth]{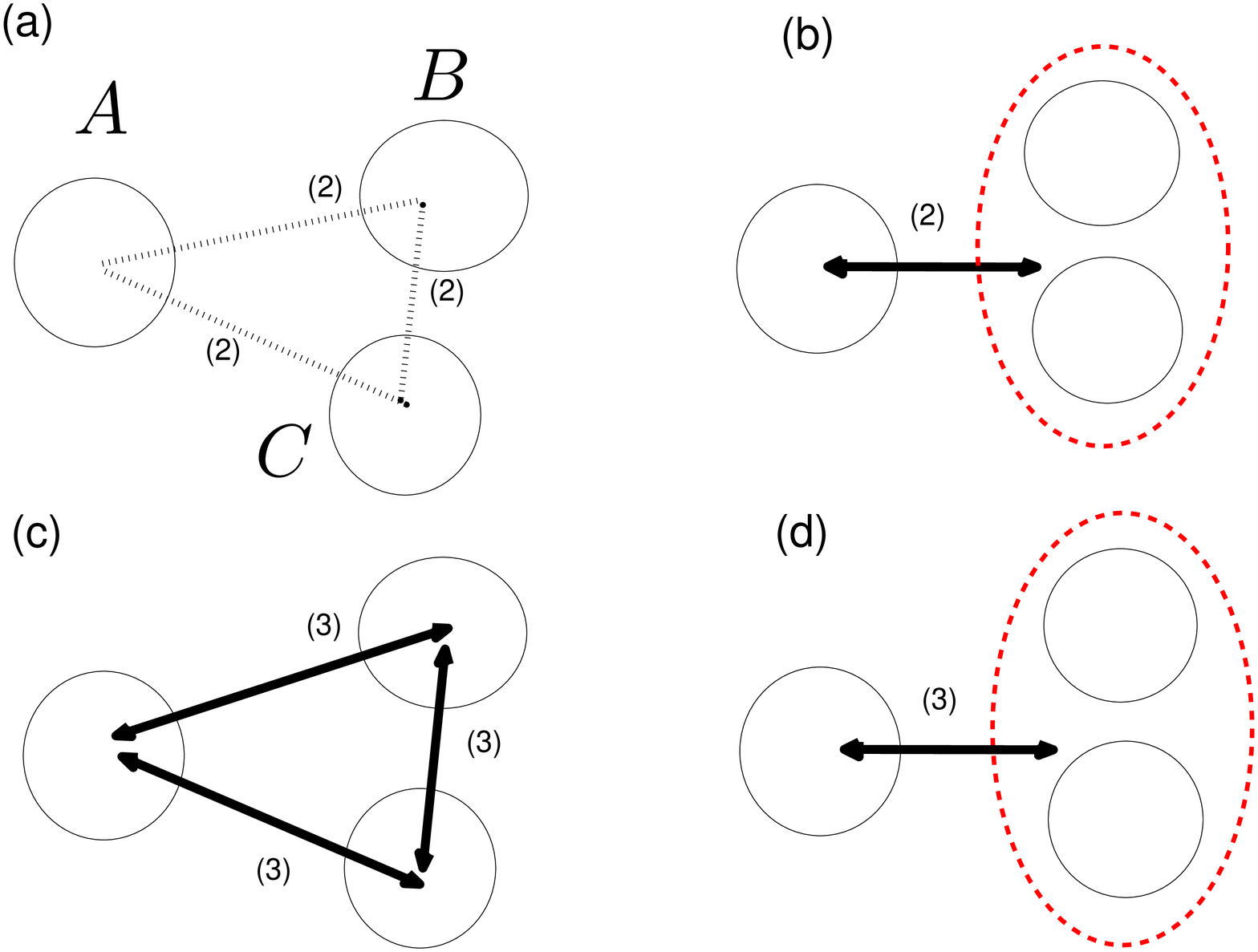}
\par\end{centering}

\caption{Composition of bipartite EPR steering for tripartite qubit $W$ and
GHZ states. (a) No bipartite EPR steering can be detected for the
W state using two-setting inequalities. Bipartite entanglement exists,
as illustrated by the dashed lines. (b) The GHZ state shows no bipartite
steering. Collective steering of one site by the group of two can
be detected using three- and two-setting inequalities. (c) Bipartite
two-way EPR steering exists for the W state, and can be detected by
the three-setting inequality.}
\end{figure}

We are now in a position to  analyse the distribution of bipartite
steering for the tripartite qubit GHZ and W states. Consider the GHZ
state, for three qubit (spin $1/2$) systems:
\begin{equation}
|\psi\rangle=\frac{1}{\sqrt{2}}\{|\uparrow\rangle_{A}|\uparrow\rangle_{B}|\uparrow\rangle_{C}-|\downarrow\rangle_{A}|\downarrow\rangle_{B}|\downarrow\rangle_{C}\}\label{eq:gh}
\end{equation}
The spins can be measured for each system, by measurements performed
by Alice, Bob and Charlie, respectively. By selecting appropriate
measurements, any two parties can predict precisely the value of any
spin component ($J^{X}$, $J^{Y}$ or $J^{Z}$) of the remaining spin
system \cite{ghz}. It was explained in Refs. \cite{he and r,EPRsteering-1}
how this implies the two- and three-observable steering of any one
party (e.g.:$B$) by the remaining group (e.g.:$AC$) i.e.: $S_{B|\{AC\}}^{(3)}=S_{B|\{AC\}}^{(2)}=0$.
It is also true that the measurement of the single spin system $B$
allows perfect inference of the orthogonal spin components of the
collective system $AC$. This implies a Bohm's EPR paradox, and hence
steering, since two spin components cannot both be specified simultaneously
in a quantum state description \cite{EPRsteering-1,bohm hans}. Such
two-way collective EPR steering for the GHZ state is depicted in Figure
5. The bipartite steering between the individual systems is evaluated,
by tracing over one system, to obtain the reduced quantum state of
the other two. As is well known \cite{entgraph,ckw}, the reduced
system is a mixture of product states, and is therefore not entangled.
Hence, there can be no bipartite steering. 

The W state \cite{W states} 
\begin{eqnarray}
|\psi\rangle & = & \frac{1}{\sqrt{3}}\{|\uparrow\rangle_{A}|\downarrow\rangle_{B}|\downarrow\rangle_{C}+|\downarrow\rangle_{A}|\uparrow\rangle_{B}|\downarrow\rangle_{C}\label{eq:w state}\\
 &  & \,\,\,\,\,\,+|\downarrow\rangle_{A}|\downarrow\rangle_{B}|\uparrow\rangle_{C}\}\nonumber 
\end{eqnarray}
gives a different sort of steering entanglement. It has been shown
that there is steering of $B$ by the group $AC$, but the steering
is reduced so that $0<S_{B|\{AC\}}^{(3)}<1$ \cite{he and r}. The
reduced state for $BA$ after tracing over $C$ is
\begin{equation}
\rho_{AB}=\frac{1}{3}\{2|\psi\rangle\langle\psi|+|\downarrow\downarrow\rangle\langle\downarrow\downarrow|\}\label{eq:mix2}
\end{equation}
where $|\psi\rangle=(|\uparrow\downarrow\rangle+|\downarrow\uparrow\rangle)/\sqrt{2}$
(we use the shortened notation $|\uparrow\uparrow\rangle\equiv|\uparrow\rangle_{A}|\uparrow\rangle_{B}$).
Conditional variances for Alice inferring Bob's results of measurement
of spin are calculated in the Appendix. If Alice measures $\sigma_{Z}^{A}$
then the average conditional variance is $(\Delta\sigma_{B|A}^{Z})^{2}=2/3$.
If she measures either $\sigma_{A}^{X}$ or $\sigma_{A}^{Y}$, then
respectively $(\Delta\sigma_{B|A}^{X})^{2}=5/9$, and $(\Delta\sigma_{B|A}^{Y})^{2}=5/9$.
Though no steering can be deduced from the two-observable inequalities
of Section II.B, the values are enough to confirm three-observable
bipartite steering since (using the expression from Section III.A)
$S_{B|A}^{(3)}\leq8/9<1$. From the symmetry of the $W$ state, we
can deduce that this steering must be two-way (Figure 5). We note
that the values are consistent with the monogamy relation (\ref{eq:smo}),
$S_{B|A}^{(3)}+S_{B|C}^{(3)}\geq S_{B|\{AC\}}^{(3)}$, that applies
in this case.

The two-observable steering behaviour is different. Here, the stricter
monogamy inequality (\ref{eq:s23}) applies: $S_{B|A}^{(2)}+S_{B|C}^{(2)}\geq2$.
For the $W$ state, we deduce that \emph{no }steering is detectable
via two-observable inequalities. The W state has complete symmetry
with respect to the three sites. Hence if there is steering of $B$
by $A$, then there must be steering of $B$ by $C$, which we have
seen is impossible for two-observable inequalities (Results 2 and
4). 
\[
\]

\section{\textbf{Discussion and Conclusion }}

The monogamy inequalities for EPR steering are likely to be useful.
For example, in order to observe EPR steering with two-setting inequalities,
we understand why it is necessary for the steering party to have greater
than 50\% efficiency for detection of data \cite{bow,buono}. Otherwise,
an eavesdropper could detect the steering also, which is forbidden
by the two-setting monogamy relation. The argument extends to the
$m$- setting inequalities, where the bound for efficiency $\eta$
is $\eta>1/m$ \cite{he and r,loopholefreesteering}.

Monogamy relations give a simple way to understand security in quantum
communication. If it can be shown that $A$ steers $B$ via a two-observable
inequality, so that $E_{B|A}$ or $S_{B|A}^{(2)}<1$, then it is guaranteed
that for a third (eavesdropper) observer $C$, $E_{B|A}$ or $S_{B|A}^{(2)}\geq1$.
Where the steering witness is directly related to the variance of
the conditional inference for Bob's values of qubits or amplitudes,
given Alice's measurements, the monogamy relations quantify the minimum
noise levels for an eavesdropper to infer Bob's values. This aids
our understanding of QKD schemes based on a shared quadrature amplitude
value, or a shared qubit value. 

The new feature associated with quantum steering is the potential
to implement \emph{one-sided device independent }cryptographic security
\cite{hw-steering-1,steer ent cry}. The noise levels for the eavesdropper
are quantified based on the uncertainty relation only, and do not
depend on the details of a particular protocol. The device-independent
security is one-way, since it is Alice's inference of Bob's amplitudes
or spin values that are secured by the steering monogamy relations.

The monogamy with respect to steering witnesses has explained the
monogamy of violations of Bell inequalities. Bell monogamy arises
because Bell inequalities are also steering inequalities. As such,
the degree of monogamy will depend on the number of observables (settings)
of the Bell inequality.

Finally, the results presented here have enabled a characterisation
of the bipartite sharing for the tripartite CV Gaussian states, and
for qubit GHZ and W states, and several experimental tests and realisations
have been proposed. Open questions remain. For example, the monogamy
results given in this paper give a quantification of how the steering
of a single system by a group is shared, but the nature of the reverse
monogamy has not been examined.
\begin{acknowledgments}
This research was supported by an Australian Research Council Discovery
grant. I thank Q He, S Armstrong, Ping Koy Lam, P Drummond, A Zeilinger
and B Wittmann for stimulating discussions on steering and related
topics.
\end{acknowledgments}

\section*{Appendix}

From (\ref{eq:mix2}), if Alice measures $\sigma_{Z}^{A}$ then the
average conditional variance is 
\begin{eqnarray*}
(\Delta\sigma_{B|A}^{Z})^{2} & = & \sum_{i}P(\sigma_{A}^{Z}=i)(\Delta(\sigma_{B}^{Z}|\sigma_{A}^{Z}))^{2}\\
 & = & \frac{1}{3}\times0+\frac{2}{3}\times1=\frac{2}{3}
\end{eqnarray*}
The joint probabilities for measurement are: $1/3$ for both Alice
and Bob with spins down; $1/3$ for Alice's spin down and Bob's up;
and $1/3$ for Alice's spin up and Bob's down. If Alice measures spin
$+1$, then Bob's state is $|\downarrow\rangle$ and the conditional
variance is $0$. If Alice measures $-1$ then Bob's spin is up and
down with probability $1/2$, and the conditional variance is $1$.
We can rewrite in the basis of spin $X$ 
\[
\rho_{AB}=\frac{1}{3}\{2|\psi_{X}\rangle\langle\psi_{X}|+|\psi_{mX}\rangle\langle\psi_{mX}|\}
\]
Here $|\psi_{X}\rangle=(|\uparrow\uparrow\rangle-|\downarrow\downarrow\rangle)/\sqrt{2}$
and $|\psi_{mX}\rangle=\frac{1}{2}\{|\uparrow\uparrow\rangle+|\downarrow\downarrow\rangle-|\uparrow\downarrow\rangle-|\downarrow\uparrow\rangle$.
If Alice measures $\sigma_{A}^{X}$ spin $+1$ (with probability $1/2$)
then the probability is $5/6$ for Bob's up and $1/6$ down, for which
the mean is $2/3$ and the conditional variance is $1-4/9=5/9$. The
same variance is obtained for outcome $-1$. Thus, $(\Delta\sigma_{B|A}^{X})^{2}=5/9$.
Rewriting in the basis of $Y$, we obtain the same conditional variance,
$(\Delta\sigma_{B|A}^{Y})^{2}=5/9$, as for spin $X$.

\end{document}